\documentclass[12pt]{article}
\input epsf.tex
\def\DESepsf(#1 width #2){\epsfxsize=#2 \epsfbox{#1}}
\usepackage{epsfig}
\usepackage{color}
\usepackage{array}
\def \gtsim    {\relax\ifmmode{\mathrel{\mathpalette\oversim >}}
                  \else{$\mathrel{\mathpalette\oversim >}$}\fi}
\def \ltsim    {\relax\ifmmode{\mathrel{\mathpalette\oversim <}}
                  \else{$\mathrel{\mathpalette\oversim <}$}\fi}
\def\oversim#1#2{\lower4pt\vbox{\baselineskip0pt \lineskip1.5pt
            \ialign{$\mathsurround=0pt#1\hfil##\hfil$\crcr#2\crcr\sim\crcr}}}
\newcommand{\invfb}{\mbox{${\rm fb}^{-1}$}}

\newcommand{\degr}{\mbox{$^{\circ}$}}
\newcommand{\Pol} {\mbox{${\cal P}$}}	
\newcommand{\Br}  {\mbox{${\cal B}$}}	
\newcommand{\pt}  {\mbox{$p_{\rm T}$}}

\newcommand{\menergy} {\mbox{${E\!\!\!\!/}$}}
\newcommand{\mpt} {\mbox{${p\!\!\!/_{\rm T}}$}}
\newcommand{\meff}{\mbox{$M_{{\rm eff}}$}}
\newcommand{\qqbar}{\mbox{$q\overline{q}$}}

\newcommand{\epem} {\mbox{$e^+e^-$}}

\newcommand{\tauh} {\mbox{$\tau_{\rm h}$}}
\newcommand{ \gluino}   {\mbox{$\tilde{g}$}}

\newcommand{ \seleRp}    {\mbox{$\tilde{e}_{R}^{+}$}}
\newcommand{ \seleRm}    {\mbox{$\tilde{e}_{R}^{-}$}}

\newcommand{ \sele}     {\mbox{$\tilde{e}$}}

\newcommand{ \smu}      {\mbox{$\tilde{\mu}$}}

\newcommand{ \stauone}  {\mbox{$\tilde{\tau}_{1}$}}
\newcommand{ \stauonep} {\mbox{$\tilde{\tau}_{1}^{+}$}}
\newcommand{ \stauonem} {\mbox{$\tilde{\tau}_{1}^{-}$}}

\newcommand{ \stau}     {\mbox{$\tilde{\tau}$}}

\newcommand{ \staup}    {\mbox{$\tilde{\tau}^{+}$}}
\newcommand{ \staum}    {\mbox{$\tilde{\tau}^{-}$}}

\newcommand{ \schionezero }{\mbox{$\tilde{\chi}_{1}^{0}$}}
\newcommand{ \schitwozero }{\mbox{$\tilde{\chi}_{2}^{0}$}}

\newcommand{ \schionepm }{\mbox{$\tilde{\chi}_{1}^{\pm}$}}

\def \etal     {\relax\ifmmode{et \; al.}\else{$et \; al.$}\fi}

\makeatletter
\def\alt{\mathrel{\mathpalette\gl@align<}}
\def\agt{\mathrel{\mathpalette\gl@align>}}
\def\gl@align#1#2{\lower.6ex\vbox{\baselineskip\z@skip\lineskip\z@
\ialign{$\m@th#1\hfil##\hfil$\crcr#2\crcr\sim\crcr}}}
\makeatother

\begin{document}
\begin{flushright}
{\tt hep-ph/0503165}\\
February, 2005 \\
\end{flushright}
\vspace*{2cm}
\begin{center}
{\baselineskip 25pt \large{\bf The Stau Neutralino Co-annihilation Region at 
an International Linear Collider} \\
}

\vspace{1cm}

{\large Vadim Khotilovich,{$^\dagger$} Richard Arnowitt,{$^\dagger$}
Bhaskar Dutta,{$^*$}
and  Teruki Kamon{{$^\dagger$}}

\vspace{.5cm}

{\it $^\dagger$Department of Physics, Texas A$\&$M University, \\
College Station, TX 77807, USA\\
$^*$Department of Physics, University of Regina, \\
Regina, Saskatchewan S4S 0A2, Canada \\

}}
\vspace{.5cm}

\vspace{1.5cm}
{\bf Abstract}
\end{center}
We probe the
stau-neutralino co-annihilation domain of the
parameter space allowed
by  the  current experimental bounds on the light Higgs mass, 
the $b\rightarrow s \gamma$ decay, and
the amount of neutralino cold dark matter  
within the framework of
minimal SUGRA models at a 500 GeV \epem\ linear collider. 
The most favorable signals of SUSY
are stau pair production and neutralino pair production
where the small mass difference between the
lighter stau and the lightest neutralino 
in the co-annihilation region is $\sim$5-15 GeV and hence
generates low-energy tau leptons in the final state. 
This small mass difference would be 
a striking signal of many SUGRA models. 
We find that a calorimeter covering down to 1\degr\ 
from the beams is crucial to reduce the two-photon background
and the mass difference could be measured at a level of 10\%
with 500 \invfb\ of data where
an invariant mass of two-tau jets and missing energy
is used as a discriminator.

\thispagestyle{empty}

\bigskip

\newpage

\section{Introduction} 
Since an international electron-positron ($e^-e^+$) linear collider (ILC) 
can measure particle masses very accurately, 
there is a growing consensus that the next high energy machine to be built 
after the Large Hadron Collider (LHC) should be an ILC. 
Such a machine is technically feasible  and 
the  initial consensus is for the TESLA design~\cite{tesla}. 
The siting  is still under discussion. 

There has been in the past a huge amount of analysis on methods of 
detecting SUSY at an ILC. 
However, the minimal supergravity (mSUGRA) model~\cite{sugra1,sugra2,nilles},  
has several special aspects that make its 
predictions clearer and  more directly accessible to experimental 
study. Hence it is worthwhile to examine this particular model.
The existing experiments have already begun to restrict the
SUSY parameter space significantly. Most significant of these are the
amount of  cold dark matter (CDM), the Higgs mass bound, 
the $b\rightarrow s \gamma$  branching ratio,
and (possibly) the muon $a_\mu$ anomaly. 

The  allowed parameter space, at present,  have three distinct regions~\cite{dark}: 
(i)~the stau neutralino ($\stau$-$\schionezero$) co-annihilation region 
where
$\schionezero$ is the lightest SUSY particle (LSP), 
(ii)~the $\schionezero$ having a 
larger Higgsino component (focus point) and 
(iii)~the scalar Higgs ($A^0$, $H^0$) annihilation funnel 
(2$M_{\schionezero}\simeq M_{A^0,H^0}$).
These three regions have been selected out by the CDM constraint. 
(There stills exists a bulk region
where none of these above properties is observed, but this region is now very small due to the existence
of other experimental bounds.) 
The distinction between the above regions can not be observed in
the dark matter experiments where only the mass of the lightest SUSY particle would be
obtained. However these regions can be observed at the ILC or the LHC where the particles will be 
 produced directly and their  masses will be measured.  

The three dark matter allowed regions need very precise measurements at the
colliders to confirm which is correct.  Since the  
ILC is  suitable for making precision measurements, 
the cosmologically allowed parameter space is under a great
deal of scrutiny.
In this paper we choose to work with the $\stau$-$\schionezero$ co-annihilation region. 
We note that many SUGRA models
possess a co-annihilation region and if the $a_{\mu}$ anomaly maintains, 
it is the only allowed region for mSUGRA.
Coannihilation is characterized by a mass difference ($\Delta M$)
between $\stau$ and $\schionezero$  of  about 5-15 GeV.
This narrow mass difference allows the $\stau$'s to co-annihilate in the early
universe along with the $\schionezero$'s in order to produce the current amount of 
dark matter density of the universe. 
The co-annihilation region has a large extension for $m_{1/2}$, up to  1-1.5 TeV,
 and can be explored at the LHC.
The main difficulty, however,  in probing this region is the small $\Delta M$ value.
This $\Delta M$ needs to be measured very accurately in order to claim that 
the  co-annihilation explains 
the dark matter content of the universe. 
However, the small $\Delta M$ value generates
signals with very low energy tau ($\tau$) leptons and thus makes  
it  difficult to discover this region at any collider due 
to the large size of the standard model (SM) and SUSY background (BG) events. 
It is this question for the ILC that we address in this paper.

At an ILC, a major source of the SM backgrounds 
is  the large two-photon  events.  
The previous studies~\cite{exp1,exp2} use
counting experiments to achieve their results.
The discovery significance is calculated using 
$N_{\rm signal}/\sqrt{ N_{\rm BG} }$ in Ref.~\cite{exp1},
while  in Ref.~\cite{exp2}, the $\stauone$ mass is measured
using the threshold method where they either scan over various
center-of-mass (CM) energies or 
assume the mass of the LSP is known from the $\sele$ and the $\smu$ decays
(to set the beam energy) in order to achieve
the maximum sensitivity for a given $\stauone$ mass.
However, as shown in Sec.~3, 
 we study the scenarios where the $\sele$ and $\smu$ masses
are too heavy to be produced at a 500 GeV machine. 
We investigate the accuracy of measuring $\Delta M$
by analyzing the shapes of 
 invariant mass distributions
of two $\tau$ jets and unbalanced event transverse momentum
(\mpt).
In our present work, we use
a fixed collider energy ($\sqrt s$ = 500 GeV) for the mass measurement.

We  first discuss the available mSUGRA parameter space in Sec.~2, 
 followed by an analysis of
 the signals and cross sections in Sec.~3.
 Monte Carlo (MC) studies on the event selection cuts
to probe the SUSY events and the SM background are reported in Sec.~4
and the precision in the mass 
measurements in Sec.~5. 
We conclude in Sec.~6.

\section{mSUGRA Parameter Space}

The models of mSUGRA depends on only four  parameters and one sign. 
These are $m_0$ (the universal soft breaking mass at the 
GUT scale $M_G$); $m_{1/2}$ (the universal gaugino soft breaking mass at $M_G$); 
$A_0$ (the universal cubic soft breaking mass at $M_G$); 
$\tan\beta = \langle H_2 \rangle / \langle H_1 \rangle$ 
at the  electroweak scale (where $H_2$ gives rise to $u$ quark masses and $H_1$ 
to $d$ quark and lepton masses); and the sign of $\mu$, the Higgs mixing 
parameter in the superpotential ($W_{\mu} = \mu H_1 H_2$). Note that the lightest 
neutralino $\schionezero$ and the gluino $\gluino$ are approximately related to 
$m_{1/2}$ by 
$M_{\schionezero} \cong 0.4\ m_{1/2}$ and $M_{\gluino} \cong 2.8\ m_{1/2}$ . 

The 
model parameters are already significantly constrained by different experimental
results. Most important 
for limiting the parameter space are:\\
\begin{itemize}
\item The light Higgs mass bound of $M_{h^0} > 114$ GeV  from LEP \cite{higgs1}. Since 
theoretical calculations of $M_{h^0}$ still have a 2-3 GeV error, we will 
conservatively assume this to mean that $(M_{h^0})^{\rm theory} > 111$ GeV.
\item The $b\rightarrow s \gamma$ branching ratio~\cite{bsgamma}. 
We assume here a relatively 
broad range (since there are theoretical errors in extracting the branching 
ratio from the data):

\begin{equation} 1.8\times10^{-4} < {\cal B}(B \rightarrow X_s \gamma) <
4.5\times10^{-4}
\label{bs}
\end{equation}

\item In mSUGRA the $\schionezero$ is the candidate for CDM. 
Previous bounds  from balloon flights (Boomerang, Maxima, 
Dasi, etc.) gave a relic density bound for CDM of $0.07 < \Omega_{\rm CDM} h^2 < 0.21 $
(where $\Omega_{\rm CDM}$ is the density of dark matter relative to the critical 
density to close the universe, and $h = H$/100 km/sec Mpc where $H$ is the Hubble 
constant). However, the new data from WMAP \cite{sp} greatly tightens this (by a 
factor of four) and the 2$\sigma$ bound is now:
\begin{equation}
 0.095 < \Omega_{\rm CDM} h^2 <0.129
\label{om}
\end{equation}

\item  The bound on the lightest chargino mass 
of $M_{\schionepm} >$ 104 GeV from LEP \cite{aleph}.

\item  The muon magnetic moment anomaly, $\delta a_\mu$, 
using both $\mu^+$ and $\mu^-$ data~\cite{BNL}. 
 Using the \epem\ data to calculate the SM leading order hadronic contribution, one gets a 2.7$\sigma$
deviation of the SM from the experimental result \cite{dav,hag}.
(The \epem\ data  appears to be more reliable than the   $\tau$ 
decay data and Conserved Vector Current (CVC)  analysis with CVC breaking~\cite{mar}.) 
Assuming the future data confirms the $a_{\mu}$ anomaly,   the combined effects 
of $g_\mu -2$ and $M_{\schionepm} >$ 104 GeV then only 
allow $\mu >0$ and leave only the $\stau$-$\schionezero$  co-annihilation 
domain of the relic density. 
\end{itemize}

One can now qualitatively state the constraints on the parameter space 
produced by the above experimental bounds:
(a)~The relic density constraint produces a narrow rising band of allowed 
parameter space in the $m_0$-$m_{1/2}$ plane;
(b)~In this band, the $M_{h^0}$ and $b\rightarrow s \gamma$ constraints produce a lower 
bound on $m_{1/2}$ for all $\tan\beta$ of 
$m_{1/2}\ \gtsim\  300\  {\rm GeV}$,                                
which implies $M_{\schionezero}> 120$ GeV and $M_{\schionepm} >$ 250 GeV. 

In the following, we will analyze the case of $\mu > 0$. 
In order to carry out the calculations it is necessary to include a 
number of corrections to obtain results of sufficient accuracy, and 
we  list some of these here: 
(i)~two loop gauge and one loop Yukawa renormalization group equations (RGEs) 
are used from $M_G$ to 
the electroweak scale, and QCD RGE below the electroweak scale for the light quarks;
(ii)~two loop 
and pole mass corrections are included in the calculation of $M_{h^0}$;
(iii)~One loop corrections to $M_b$ and $M_\tau$  are included~\cite{rattazi};
(iv)~large $\tan\beta$ SUSY corrections to $b\rightarrow s \gamma$ are included
\cite{degrassi};
(v)~all  $\stauone$-$\schionezero$ co-annihilation channels are included in the 
relic density calculation \cite{bdutta}. 
We do not include Yukawa unification or proton decay constraints as these 
depend sensitively on post GUT physics, about which  little is known.

\begin{figure}[htb]\vspace{0cm}
\hspace*{0.0cm}
\centerline{
\epsfxsize=8.6cm\epsfysize=9cm
\epsfbox{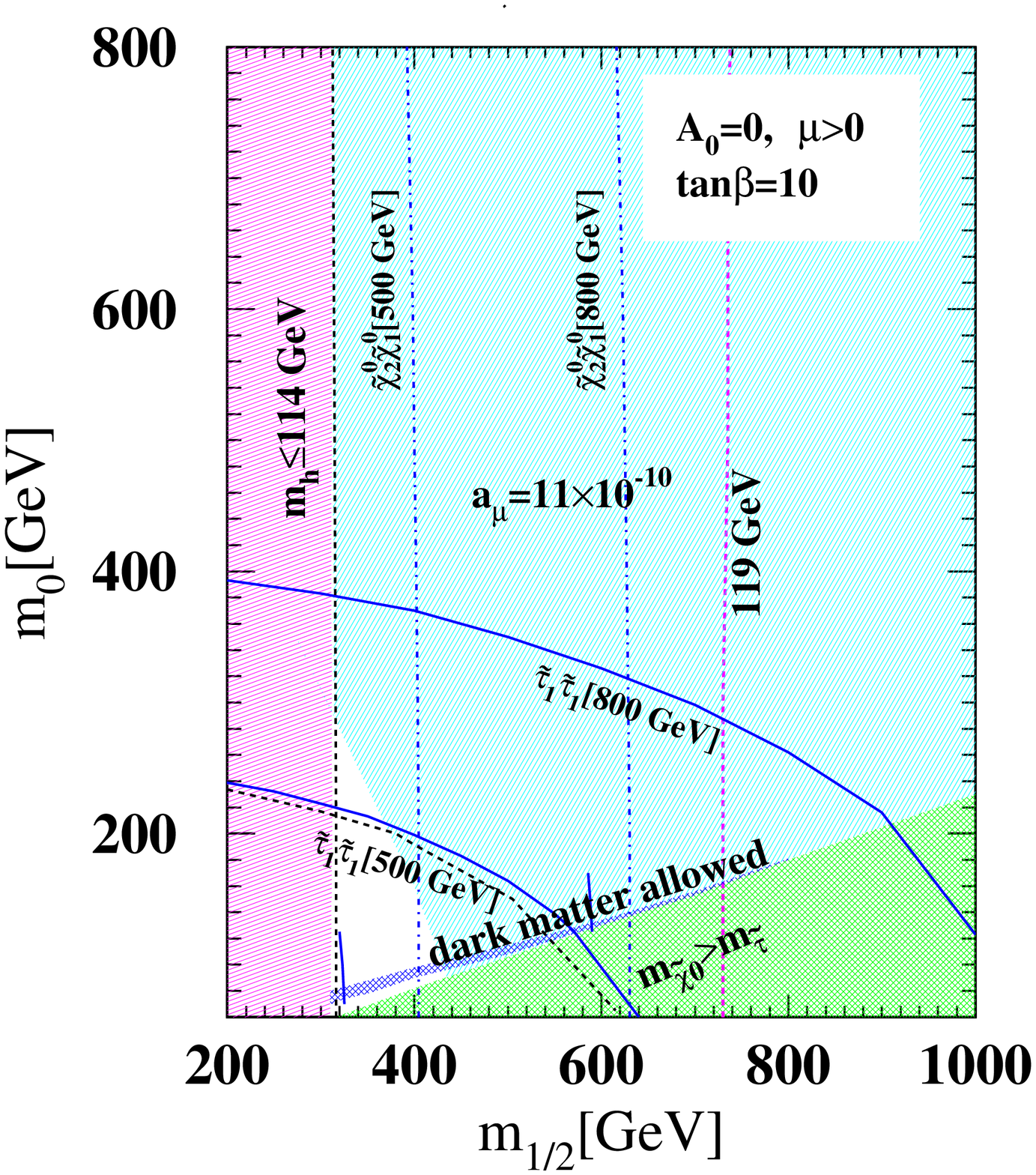}}
\vspace*{0.5cm}

\centerline{\epsfxsize=8.6cm\epsfysize=9cm
\epsfbox{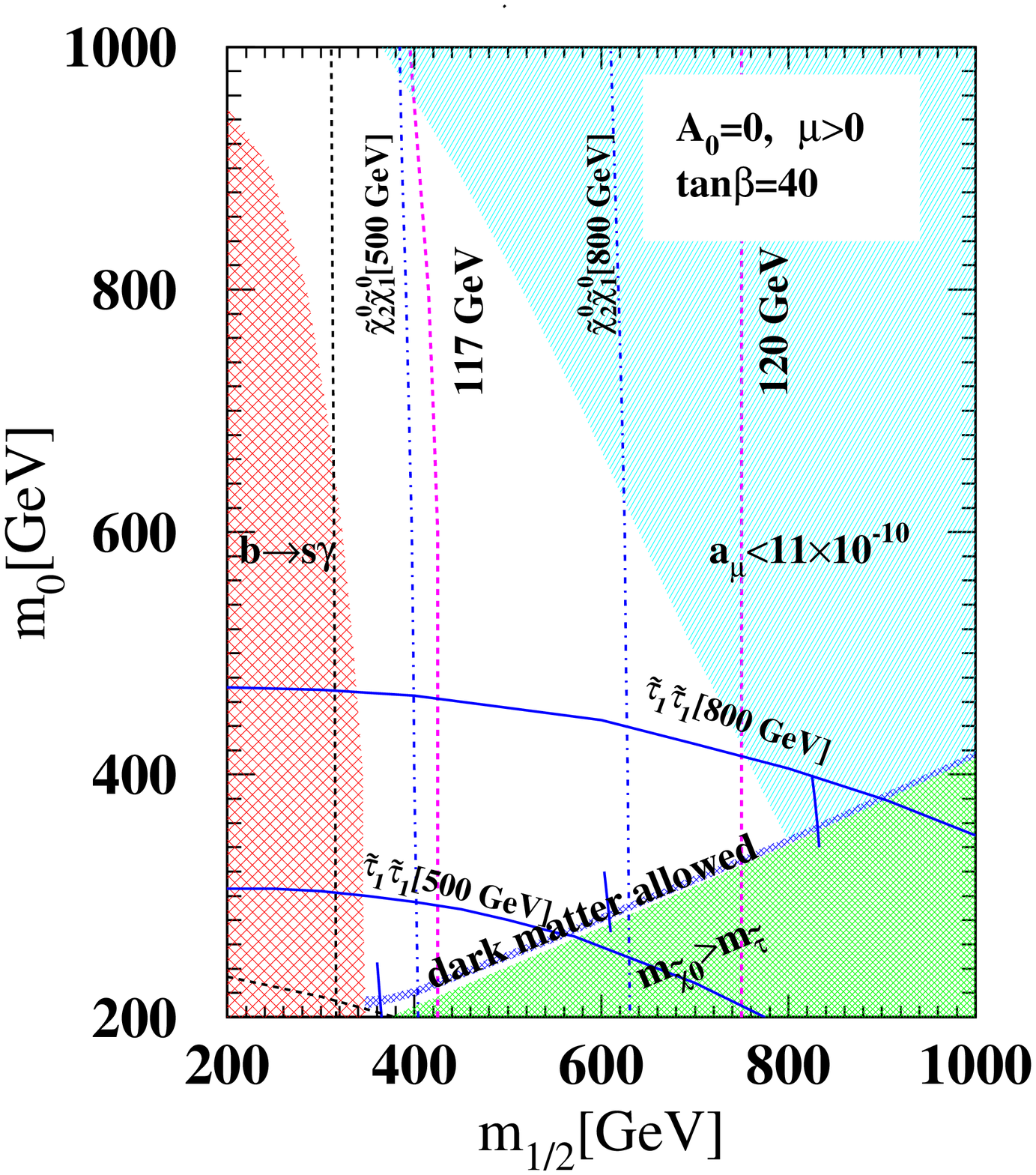}
\hspace*{0.2cm}
\epsfxsize=8.6cm\epsfysize=9cm
\epsfbox{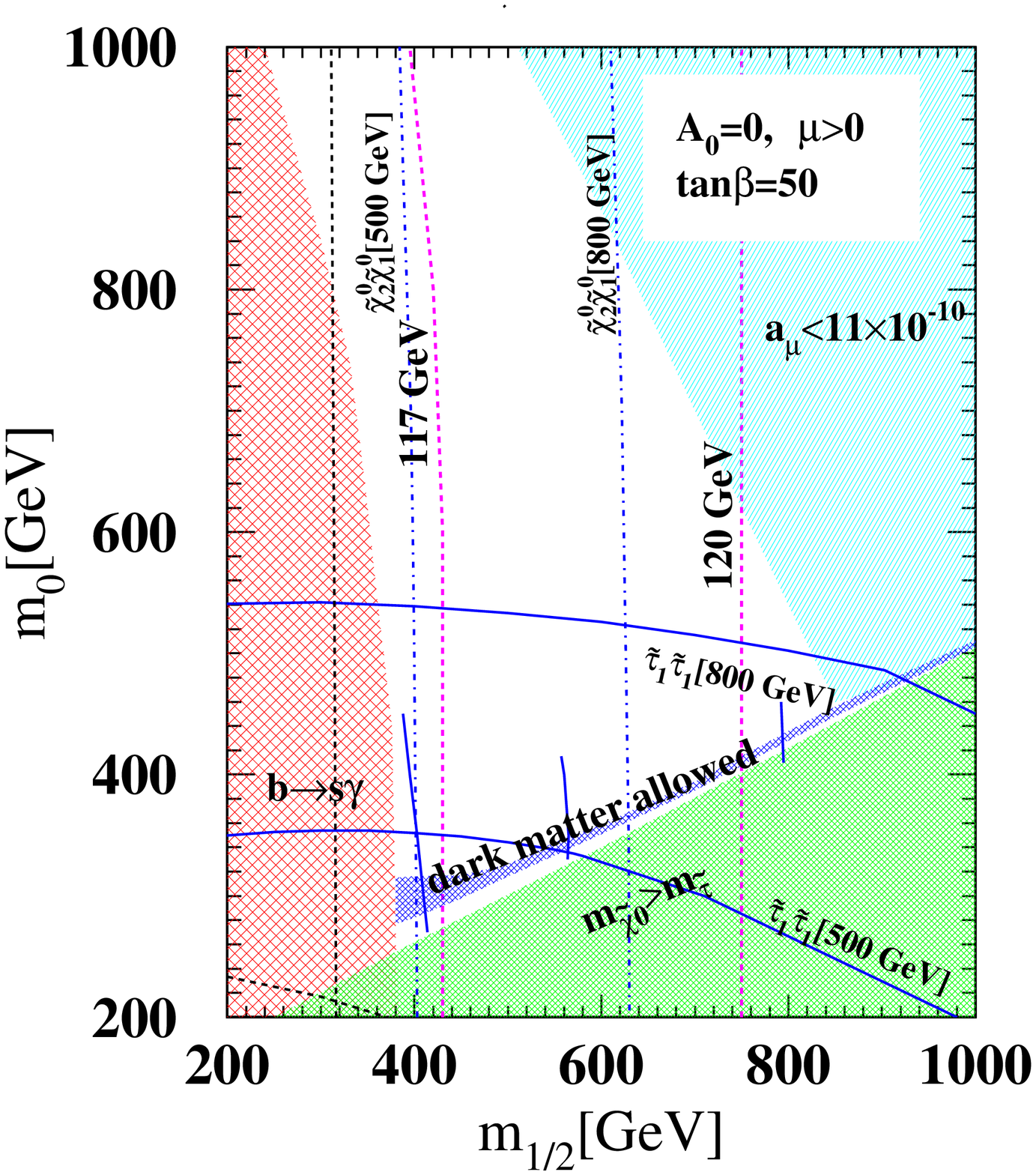}
\vspace*{0.0cm}}
\caption[fig:fig2]{
Allowed region in the $m_0$-$m_{1/2}$ plane from the relic density 
constraint  for $\tan\beta$ = 10, 40, and 50
with  $A_0  = 0$ and $\mu >0$. 
The narrow blue band by the WMAP data. 
The dotted pink vertical lines are different Higgs masses, and the 
current LEP bound produces a lower bound on $m_{1/2}$ for low $\tan\beta$.
The brick red region depicts the 
$b\rightarrow s \gamma$ constraint for
$\tan\beta=40$ and 50. 
For $\tan\beta=10$, the pink region depicts the Higgs mass
region $M_{h^0} \leq$ 114 GeV.  The light blue region 
is excluded if $\delta a_{\mu} > 11 \times 10^{-10}$.
(Other lines are discussed in text.)} 
\label{WMAP_allowed_region}
\end{figure}

Figure~\ref{WMAP_allowed_region} illustrates the  constraints on the mSUGRA parameter 
space for $\tan\beta$ = 10, 40 and 50  with $A_0$ = 0. 
The narrow blue band is the region now allowed by WMAP (see Eq.~\ref{om}). 
The dotted 
pink lines are for different Higgs masses, and the light blue region would 
be excluded if $\delta a_\mu > 11\times10^{-10}$. The three short solid lines 
indicate the $\schionezero$-$p$ cross section values. 
In the case of $\tan\beta=40$ they represent (from left) 0.03 $\times 10^{-6}$ pb,  
0.002 $\times 10^{-6}$ pb, 0.001 $\times
10^{-6}$ pb and in the case of $\tan\beta=50$ they represent 
(from left) 0.05 $\times 10^{-6}$ pb,  0.004 $\times 10^{-6}$ pb, 
0.002 $\times 10^{-6}$ pb. In the case of $\tan\beta=10$ they represent 
(from left) 5 $\times 10^{-9}$ pb and 1 $\times
10^{-9}$ pb.
It is important to note that the 
narrowness of the allowed dark matter band is not a fine tuning. The lower 
limit of the band comes from the rapid annihilation of neutralinos in the 
early universe due to co-annihilation effects as the light $\stauone$ mass, 
$M_{\stauone}$, approaches the neutralino mass as one lowers $m_0$. 
Thus the lower 
edge of the band corresponds to the lower 
bound of Eq.~\ref{om}, and the band is 
cut off sharply due to the Boltzman exponential behavior. 
The upper limit 
of the band, corresponding to the upper bound of Eq.~\ref{om},
 arises due to 
insufficient annihilation as $m_0$ is raised. As the WMAP data becomes more 
accurate, the allowed band will narrow even more. (Note that the 
slope and position of the band changes, however as $A_0$ is changed.) 
Thus the 
astronomical determination of the amount of dark matter  effectively 
determines one combination of the four parameters of mSUGRA. 
Since  the $\stau$-$\schionezero$ 
co-annihilation region seems to be experimentally most favored (including the $g-2$
effect), we probe this region. 
Let us now study the available sparticles when we
try to probe this co-annihilation band in a linear collider.

\section{Production and Signals of SUSY Particles at an ILC}

Figure~\ref{WMAP_allowed_region}  shows
the production cross section of 0.1 fb
for $\seleRp\seleRm$ (black dashed),
$\stauonep\stauonem$ (blue solid), $\schionezero\schitwozero$ (blue
dashed-dot) and chargino pair (vertical black dot)
productions.  
We see that for large $\tan\beta$ the chargino pair production 
is almost not observable and the selectron pair
production is unobservable.
The stau pair has the largest reach in $m_{1/2}$ and the neutralino pair
has the largest reach in $m_0$. 
We therefore focus on the stau pair and the neutralino pair production
cross sections. 
The  kinematical reach of the production cross sections 
of $\stauonep\stauonem$ and 
$\schionezero\schitwozero$ productions
for both $\sqrt s=$ 500 and 800 GeV are shown in the figures. 
We see that the 800 GeV ILC will have a much bigger reach. 
 We will, however, use the 500 GeV collider to study the signal 
 since it seems to be the
 intial CM energy for  ILC.
 
  The possible signals   for $\stauonep\stauonem$ and $\schionezero\schitwozero$ in 
mSUGRA are the following:
\begin{eqnarray} 
\epem & \rightarrow & \stauonep \stauonem \rightarrow 
( \tau^+ \schionezero) + ( \tau^- \schionezero)\\
\epem & \rightarrow & \schionezero\ \schitwozero  
\rightarrow \schionezero + (\tau \stauone) \rightarrow  
	    \schionezero + (\tau^+ \tau^- \schionezero)
\end{eqnarray}
We look at the hadronic final state of taus (\tauh's)
in order to have larger event rates. 
The final signal thus 
has  two $\tauh$'s plus \mpt.
The analysis now is quite complicated  
since the $\tau$'s have low energy due to a small $\Delta M$ value.
We need to develop appropriate event selection cuts to extract
the signal from the SM background which is dominated 
by the $\gamma^* \gamma^*  e^+ e^-$.

In general, the co-annihilation region occurs for $\Delta M\,\sim$
5-15 GeV. 
We choose three points for $m_{1/2}$ = 360 GeV, $m_0$ = 205, 210 and 220 GeV, 
with $A_0$ = 0  and $\tan\beta=40$ and develop our event selection cuts. 
The masses of SUSY particles in these three representative scenarios 
are given in Table \ref{table:Msusy}.
The values of $\Delta M$ for these three points are 5, 10 and 19 GeV. 
The first selection we use is the electron beam polarization. 
Since both signals and background processes cross sections are affected by it,
we choose appropriate polarization to increase the significance of the signal. 
The production cross sections  for $\sqrt s$ = 500 GeV for different polarizations 
are given in Table~\ref{table:Xsec}. 
The right-handed (RH) polarization $\Pol(e^-) = -0.9$  enhances 
the $\stauonep \stauonem$ signal, 
and the left-handed (LH) polarization, 
$\Pol(e^-) = +0.9$ enhances the $\schionezero \schitwozero$ signal.
The SM background, mentioned in table, consists of $\bar\nu \nu \tau^+ \tau^-$ 
states arising from $WW$, $ZZ$ and $Z\nu\nu$
production and this background becomes smaller for
a right-handed electron beam. 
In addition to this, we also have two photon processes which will
be described  later:
$e^+ e^- \rightarrow \gamma^* \gamma^* +  e^+ e^-\rightarrow \tau^+ \tau^-$ 
(or $q \bar{q}$) + $e^+ e^-$ 
where the final state $e^+ e^-$ pair are at a  small angle to the beam pipe and 
the $\qqbar$ jets fake a $\tau^+ \tau^-$ pair. This background, does not change with beam
polarization and needs to be suppressed  by appropriate cuts.

\begin{table}[ht]
\caption{Masses (in GeV) of SUSY particles in three representative scenarios of 
$\Delta M \equiv M_{\stauone} - M_{\schionezero}$
for $m_{1/2}$ = 360 GeV, $\tan\beta = 40$, $\mu > 0$, and $A_0 = 0$. 
These points satisfy all the existing experimental bounds on mSUGRA.
The numbers were obtained 
using {\tt ISAJET} \protect\cite{isajet}.}
\label{table:Msusy}
\begin{center}
\begin{tabular}{l c c c c }
\hline \hline  MC Point & $M_{\schitwozero}$ &  $M_{\stauone}$ & 
$M_{\schionezero}$ &
        $\Delta M$  \\
($m_0$ in GeV) & & & &  \\
\hline
 1 (205) &  274.2 &  147.2 &  142.5 &  4.76   \\
 2 (210) &  274.2 &  152.0 &  142.5 &  9.53   \\
 3 (220) &  274.3 &  161.6 &  142.6 &  19.0   \\
\hline \hline
\end{tabular}
\end{center}
\end{table}

\begin{table}[ht]
\caption{Cross section times branching ratio (in fb),
$\sigma \times \Br(\tau \rightarrow \tauh)^{2}$, for
SUSY and SM 4-fermions (4f) production  in two cases of 
polarizations,  $\Pol(e^-) = -0.9$ (RH) and +0.9 (LH).
The SUSY cross sections were obtained using {\tt ISAJET}~\protect\cite{isajet},
and {\tt WPHACT}~\cite{wphact} was used for the cross sections of the $\bar\nu \nu \tau^+ \tau^-$ processes.}
\label{table:Xsec}
\begin{center} 
\begin{tabular}{l c | c  c }
\hline \hline 
 & $\Pol(e^-)$ & $-$0.9 (RH)& 0.9 (LH)\\\hline
SM 4f         &                            & 7.84 & 89.8 \\
\hline
SUSY point 1 & $\schionezero\schitwozero$ & 0.41  & 6.09 \\
                    & $\stauonep\stauonem$ & 28.3 & 13.2 \\
	\hline
SUSY point 2 & $\schionezero\schitwozero$ & 0.40  & 6.00 \\
                    & $\stauonep\stauonem$ & 26.6 & 12.4 \\
\hline
SUSY point 3 & $\schionezero\schitwozero$ & 0.38  & 5.68 \\
                    & $\stauonep\stauonem$ & 23.0 & 10.6 \\
\hline \hline
\end{tabular}
\end{center}
\end{table}

  The event selection cuts with the LH polarization will be  
optimized to enhance the $\schionezero\schitwozero$ signal 
and the RH cuts to optimize the 
$\stauonep\stauonem$ signal.   

The generation of MC samples and the analysis for the signal and the background 
was done using the following programs: 
(1)~{\tt ISAJET}~\cite{isajet} to generate SUSY events; 
(2)~{\tt WPHACT}~\cite{wphact} for SM backgrounds;
(3)~{\tt TAUOLA}~\cite{tauola} for tau decay; 
(4)~Events were simulated and analysed with a LC detector simulation~\cite{LCD}.

\subsection{Event Selection}

\begin{table}[t]
\caption{Kinematic cuts 
for the LH ($\Pol = 0.9$) and the RH ($\Pol = -0.9$) cases}
\label{table:EventSelectionCuts}
\begin{center}
\begin{tabular}{ l | c c c }
\hline \hline
Cut Variable(s) & LH ($\Pol(e^-)$ = 0.9) & ~~ & RH ($\Pol(e^-)$ = $-$0.9) \\
\hline \hline  
$N_{jet}$($E_{jet}>3$ GeV) & \multicolumn{3}{c}{2} \\
$\tau_h$ ID & \multicolumn{3}{c}{1, 3 tracks; $M_{tracks} <$ 1.8 GeV} \\
\hline
Jet acceptance & $-q_{jet} \cos\theta_{jet} < 0.7$ & ~~ &  $|\cos\theta_{jet} | < 0.65$\\
 & $-0.8 < \cos\theta(j_2,p_{vis})  < 0.7$ & ~~ & $| \cos\theta(j_2,p_{vis}) | < 0.6$\\\hline
Missing \pt & \multicolumn{3}{c}{$> 5$ GeV}\\ \hline
Acoplanarity & \multicolumn{3}{c}{$> 40\degr$} \\
\hline
Veto on EM clusters & \multicolumn{3}{c}{No EM cluster in $5.8\degr < \theta < 28\degr$ with $E > 2$ GeV} \\ 
 or electrons & \multicolumn{3}{c}{No electrons within $\theta > 28\degr$ with $\pt > 1.5$ GeV} \\
 \hline
Very forward calorimeter (1\degr (2\degr) - 5.8\degr) &
 \multicolumn{3}{c}{No EM cluster with $E>100$ GeV} \\
\hline\hline
\end{tabular}
\end{center}
\end{table}

In order to reduce the backgrounds we require
a set of event selection cuts and 
these cuts are given in Table~\ref{table:EventSelectionCuts}. 
In this table: $j_2$ stands for second leading $\tau$ jet,
$p_{vis}$ gives the sum of  visible momenta and $\theta(j_2,p_{vis})$ is the angle
between them. $\theta_{jet}$ is the angle between a $\tau$  jet and the beam direction.
The jets are reconstructed using the JADE algorithm with $Y_{{\rm cut}} \geq 0.0025$~\cite{jade} 
and selected with $E_{jet} >$ 3 GeV.
Such a value of the  $Y_{{\rm cut}}$ parameter helps to select narrow $\tau$-like jets.
The jet acceptance cut is required to reduce the SM background events such as $WW$ and $ZZ$ production.
The acoplanarity is defined as ${\cal A} = 180\degr - \Delta\phi(j_1,j_2)$, 
where $\Delta\phi(j_1,j_2)$ is the azimuthal angle 
between two $\tau$-jets. 
The cut on acoplanarity is very powerful in rejecting two photon SM backgrounds which
have a huge cross section.
In order to have MC samples of manageable size for two photon SM processes 
we apply a cut ${\cal A}_{MC} > 30\degr$ already at the  generator level. 
(In addition for these samples we apply the generator level cut on $\pt^{\tau MC} > 4$ GeV 
and require the $\tau$ to be separated from the beam line by more than 35\degr).
We also require no EM clusters (a)~in $5.8\degr < \theta < 28\degr$ where the ILC detector
has no tracking system and 
(b)~in the angle below $5.8\degr$ with two options of a very forward calrometer
(VFD). 
In our calculation, beamstrahlung and bremsstrahlung are
included in the two-photon annihilation process. 
The two photon background in our analysis is similar to that  discussed in Refs.~\cite{exp1,exp2}. 

The number of accepted events for each class of final states for the case \mpt\, $>$ 5 , 10, and 20 
GeV are summarized in Table~\ref{table:Nevent_500invfb}.

\begin{itemize}
\item The RH polarization strongly suppresses the  SM background events
($WW$ etc.) and 
the neutralino events ($\schionezero\schitwozero$ ). We also need    
 a 1\degr\ VFD and \mpt\ $>$ 5 GeV to get  a clean 
signal for the $\stauonep\stauonem$ events. 
With no VFD there would be approximately 4,400
SM $\gamma\gamma$ background events swamping the SUSY signal.

\item The LH polarization  allows for 
the detection of the $\schionezero\schitwozero$ signal
with \mpt\ $>$ 20 GeV without a VFD 
(as the $\gamma\gamma$ background falls to zero then),
or \mpt\ $>$ 10 GeV with a 2\degr\ VFD. 
However, both a 1\degr\ VFD and \mpt\ $>$ 5 GeV are necessary
to detect the $\stauonep\stauonem$ events and to measure
$\Delta M$  in the LH case.
In the case of no VFD there would be $\sim$9,300
SM $\gamma\gamma$ background events with \mpt\ $>$ 5 GeV. 
Note that the event selection criteria in
the LH polarization case are different from the RH case.
\end{itemize}

Thus we find that  the VFD is 
essential to detect SUSY in this region of parameter space. 
A lower \mpt\ increases the number of 
events and the significance. 
A 5 GeV \mpt\ cut has been found to be feasible at a 500 GeV ILC.

It should be noted that the
1\degr\ VFD is feasible for the ILC
 since  the TESLA design (which has been accepted for
 the ILC technology) allows a VFD coverage
down to 3.2 mrad (or 0.18$^\circ$)~\cite{tesla}.
We also note that
our study is based on head-on collisions of electron and positron.
However,  it has been shown that the VFD  is still able to
   reduce the two-photon background events
    even in the case of a  beam  crossing~\cite{exp2}. 

\begin{figure}[htb]\vspace{-4cm}
\centerline{
\epsfxsize=12.6cm\epsfysize=16.0cm
\epsfbox{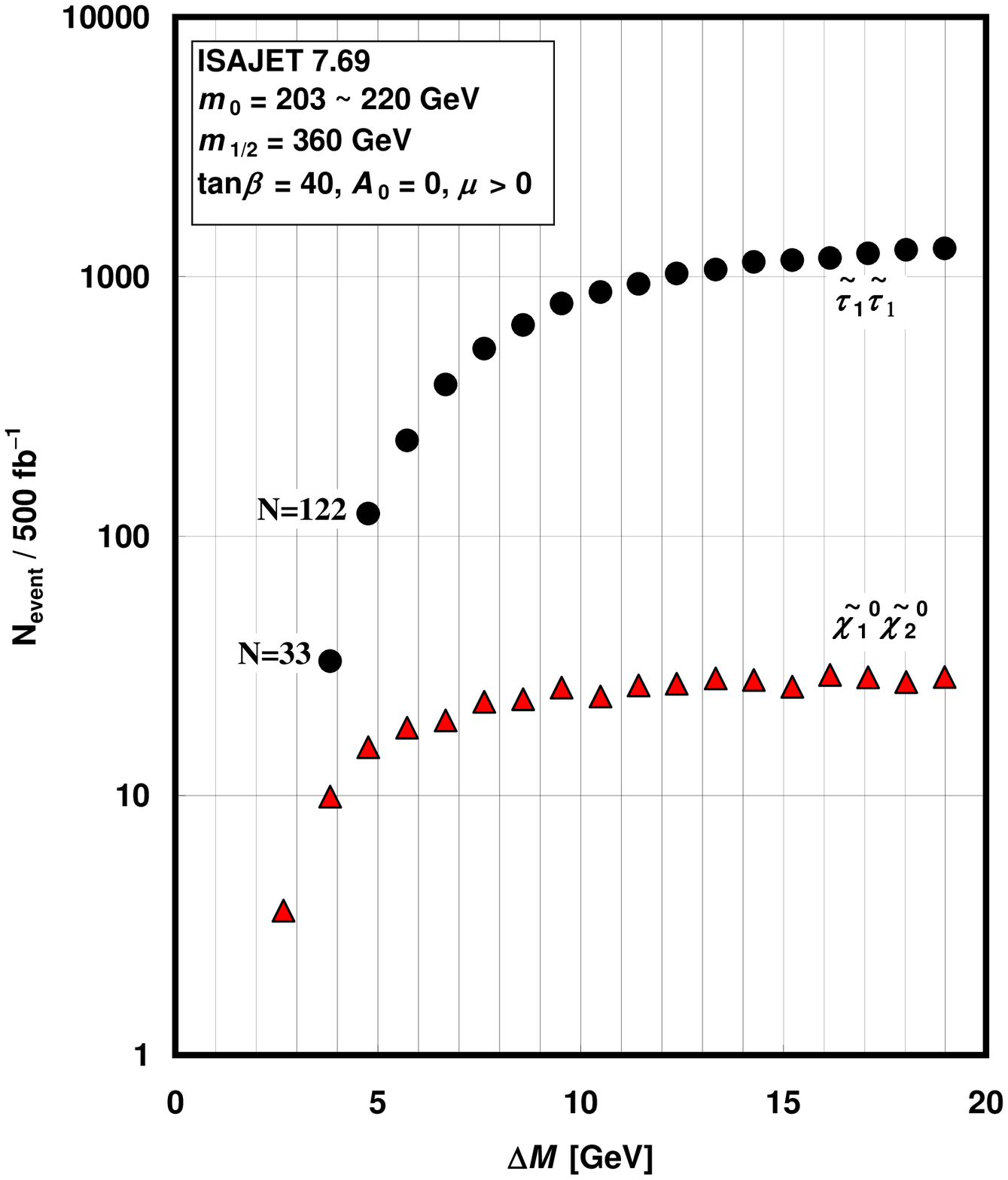}}\vspace{-2cm}
\caption[fig:fig5]{Number of $\tau_h\tau_h\schionezero\schionezero$ events
from $\stauonep\stauonem$ (solid circles)
and $\schionezero\schitwozero$ (solid triangles) production 
as a function of $\Delta M$ (for $m_0$ = 203-220 GeV at $m_{1/2}$ = 360 GeV)
in the RH polarization case. 
We assume 500 \invfb\ of luminosity.} 
\label{fig5}
\end{figure}

\begin{figure}[htb]
\centerline{
\epsfxsize=7.6cm\epsfysize=9.0cm
\epsfbox{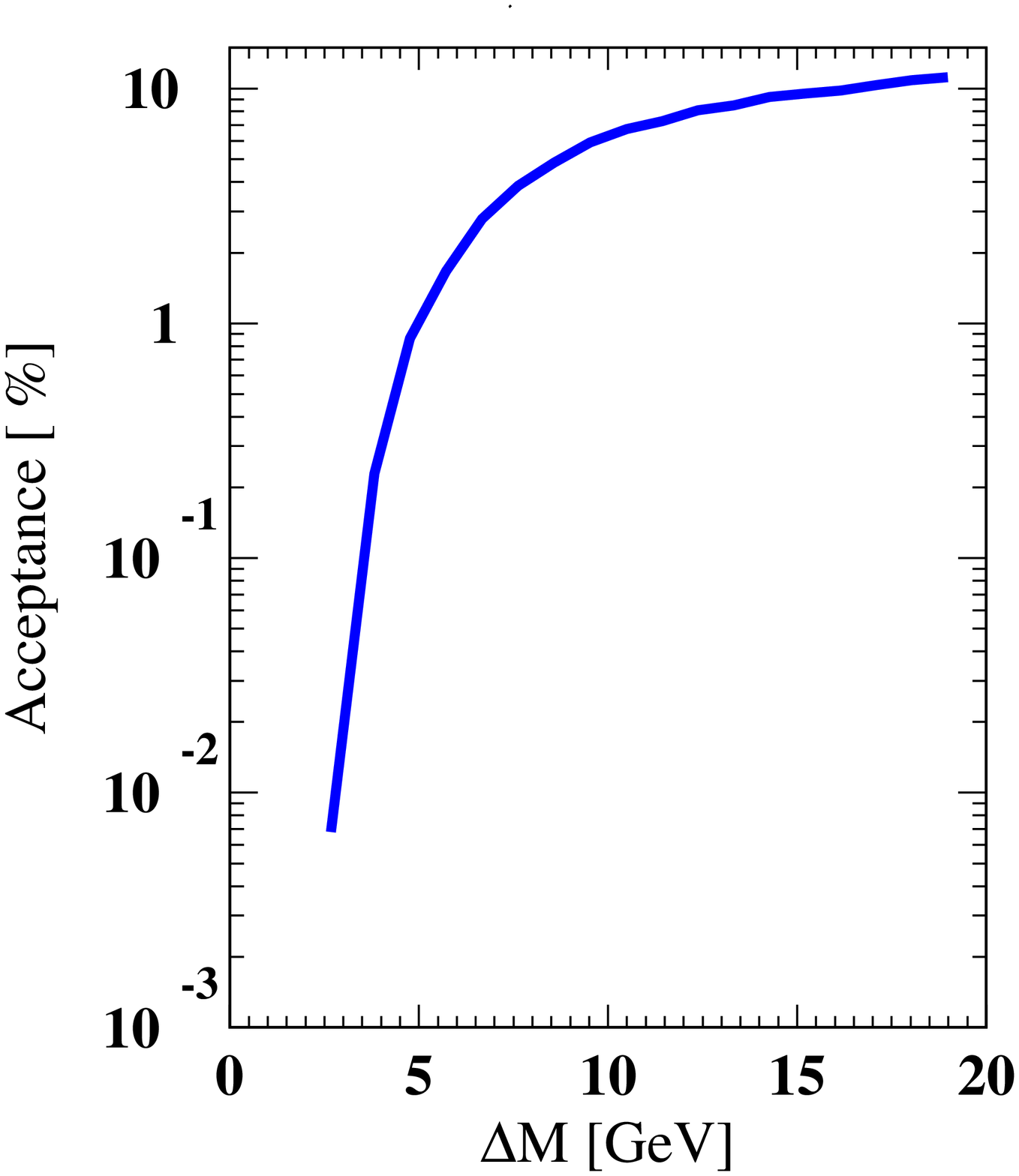}}
\caption[fig:fig5b]{Total event acceptance for 
$\stauonep\stauonem \to \tauh\tauh\schionezero\schionezero$
as a function of $\Delta M$ for $m_0$ = 203-220 GeV 
($m_{1/2}$ = 360 GeV) in the RH polarization case. } 
\label{fig5B}
\end{figure}

The $\stauonep\stauonem$ cross section has the largest reach 
along the co-annihilation band and one would use this channel 
to measure the mass difference. 
This channel needs RH polarization for enhancement.
In Figure~\ref{fig5}, we plot the number of events accepted
in our selection criteria with 500 \invfb\
of luminosity as a 
function of $\Delta M$ for $m_0 =$ 203-220 GeV
($m_{1/2}$ = 360 GeV) in the RH polarization case. 
We see that we have  more than 100 events, which 
will be adequate for the measurement of $\Delta M$
as discussed in Section \ref{sec:MassMeasurement},
for $\Delta M > 4.5$ GeV. 
Figure~\ref{fig5B} is a plot of the acceptance 
as a function of $\Delta M$ for $m_0 =$ 203-220 GeV 
with a $1\degr$ VFD  in the case of RH polarization. 
The acceptance drops rapidly as $\Delta M$ goes below 5 GeV.  

\begin{figure}[htb]
\centerline{
\epsfxsize=10cm\epsfysize=10.0cm
\epsfbox{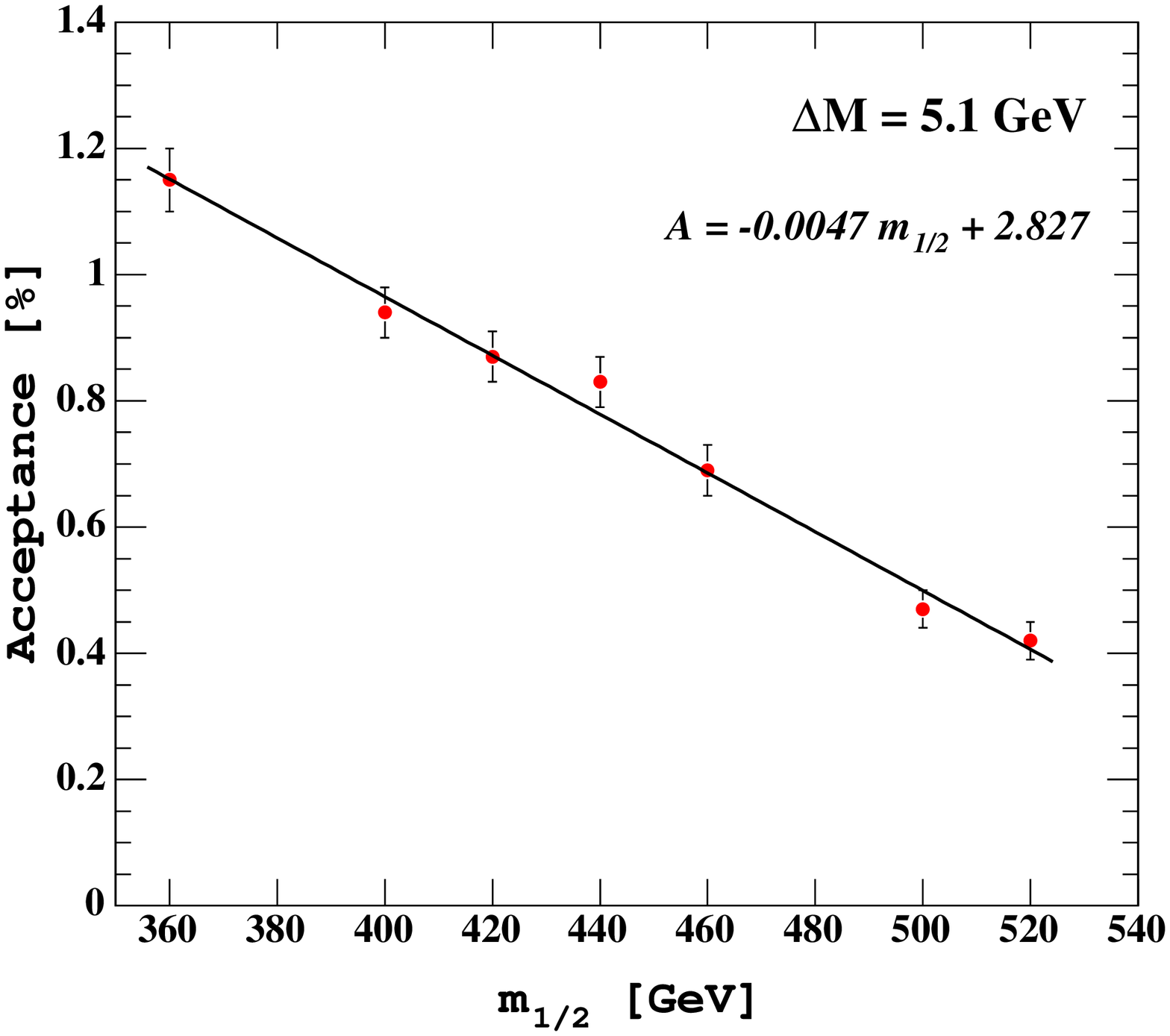}}
\caption[fig:fig5x]{Total event acceptance for 
$\stauonep\stauonem \to \tauh\tauh\schionezero\schionezero$
as a function of $m_{1/2}$ for $\Delta M$ = 5.1 GeV in the RH polarization case. } 
\label{figaccpt}
\end{figure}

The event acceptance also depends on $m_{1/2}$ 
as shown in Figure~\ref{figaccpt}.
This dependence arises because the $\tau$'s are less energetic
and its angular distribution changes as the stau becomes heavier.
We calculate the significance ($\sigma$) as
$N_{{\rm signal}}/ \sqrt{ N_{{\rm BG}} }$,
where the  
$\schionezero\schitwozero$ events are also treated as backgrounds, for a window of 
 $\meff\
\equiv M(j_1,j_2,\menergy)$ (invariant mass of two $\tau$-jets
and missing energy). 
For $\Delta M$ = 4.76 and 19 GeV, 
the allowed ranges for \meff\ are 0-54.5 GeV and 0-183.5 GeV, respectively.
The 5$\sigma$ reach for the $\stauone$ mass is found to be  
$\leq$ 215 GeV ($m_{1/2}\leq$  520 GeV) 
for $\Delta M$ = 4.76 GeV  with a 1\degr\ VFD and  \mpt\ $>$ 5 GeV. 
For $\Delta M$ = 19 GeV,
the $5\sigma$ reach of the $\stau$ mass at a 500 GeV ILC 
is $\leq$ 237 GeV ($m_{1/2}\leq$  537 GeV).

It should be noted that our event selection cuts  
are optimized for a 500 GeV machine.
In the case of an 800 GeV ILC, the cuts need to be re-optimized
based on the new SUSY backgrounds
and machine design limitations (e.g. the lower bound on \mpt\ needs to be raised).

\begin{table}[t]
\caption{Number of $\tau_h\tau_h$ plus \mpt\ events for luminosity of 500 fb$^{-1}$  for 
points 1, 2 and 3 corresponding to $\Delta M$ = 4.76, 9.5, and 19.0 GeV,
respectively. 
All numbers except for two-photon backgrounds are common
for different options of VFD.
\label{table:Nevent_500invfb}}
\begin{center}
\begin{tabular}{l l | r r r c |c r r r }
\hline
\hline
  & &         \multicolumn{4}{c|}{${\cal P}(e^-)=0.9$ (LH)} &
        \multicolumn{4}{c}{${\cal P}(e^-)=-0.9$ (RH)} \\
Process &    & $\mpt^{min}$ = 5  & 10 & 20 & ~ & ~ & 5 & 10 & 20 \\
\hline
$\schitwozero \schionezero$
        & Pt.1  & 374 & 342 & 260 & ~ & ~ & 15 & 14 & 11 \\
        & Pt.2  & 624 & 572 & 425 & ~ & ~ & 26 & 24 & 18 \\
        & Pt.3  & 743 & 697 & 529 & ~ & ~ & 29 & 28 & 22 \\
\hline
$\staup \staum$
        & Pt.1  & 73 & 2 & 0 & ~ & ~ & 122 & 2 & 0 \\
        & Pt.2  & 524 & 267 & 11 & ~ & ~ & 786 & 437 & 22 \\
        & Pt.3  & 946 & 781 & 335 & ~ & ~ & 1283 & 1076 & 468 \\
\hline
SM 4f     &      & 1745 & 1626 & 1240 & ~ & ~ & 129 & 123 & 100 \\
SM $\gamma\gamma$ & 2-5.8$^\circ$ VFD  & 535 & 7 & 0 & ~ & ~& 249 & 4 & 0\\
             & 1-5.8$^\circ$ VFD & 10 & 0 & 0 & ~ & ~ & 4 & 0 & 0\\
\hline
\hline
\end{tabular}
\end{center}
\end{table}

\section{Measurement of Stau Neutralino Mass Difference}
\label{sec:MassMeasurement}

\begin{figure}\vspace{-2cm}
\centerline{\epsfxsize=10cm\epsfysize=10.0cm
\epsfbox{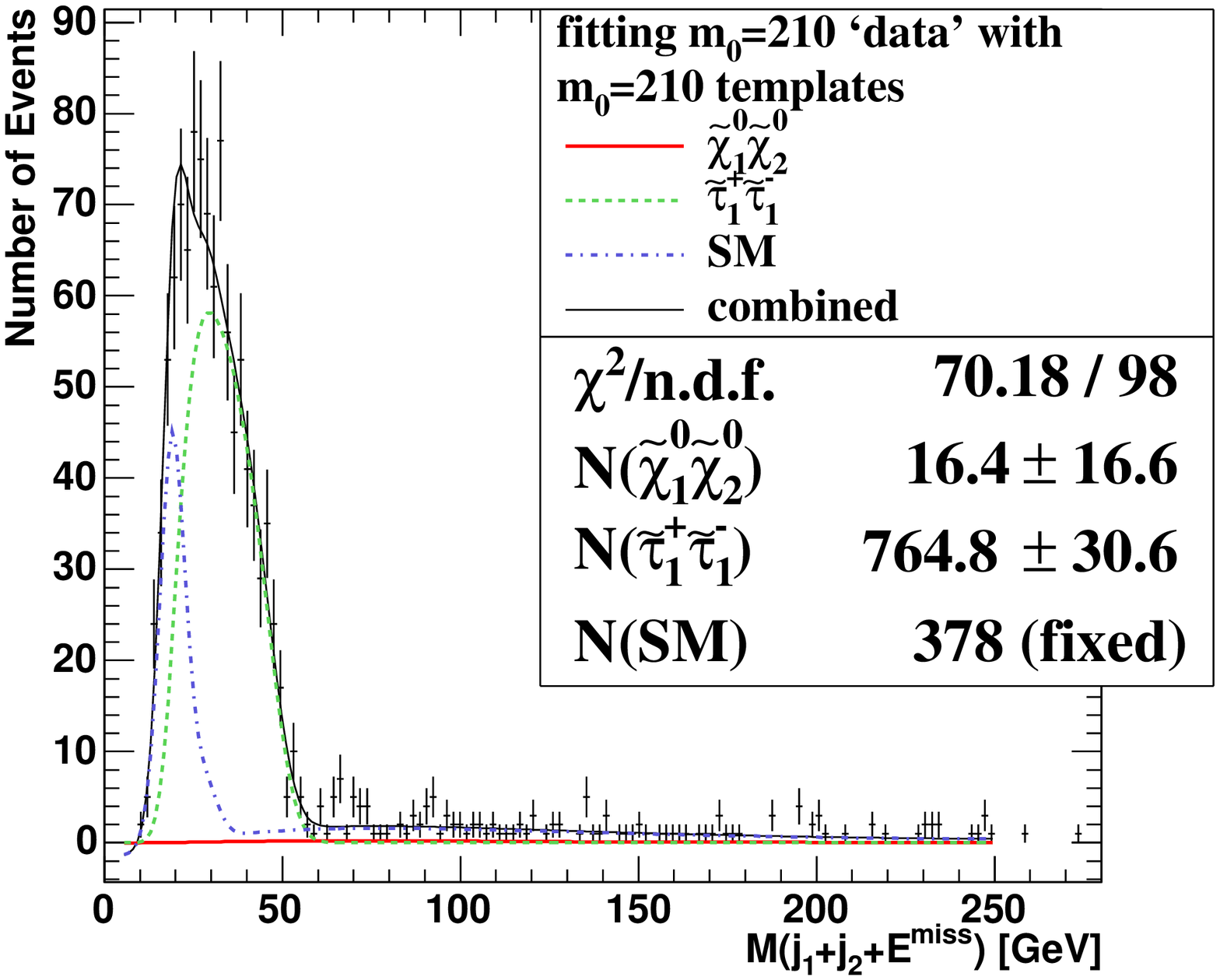}}\vspace*{-0cm}
\centerline{\epsfxsize=10cm\epsfysize=10.0cm
\epsfbox{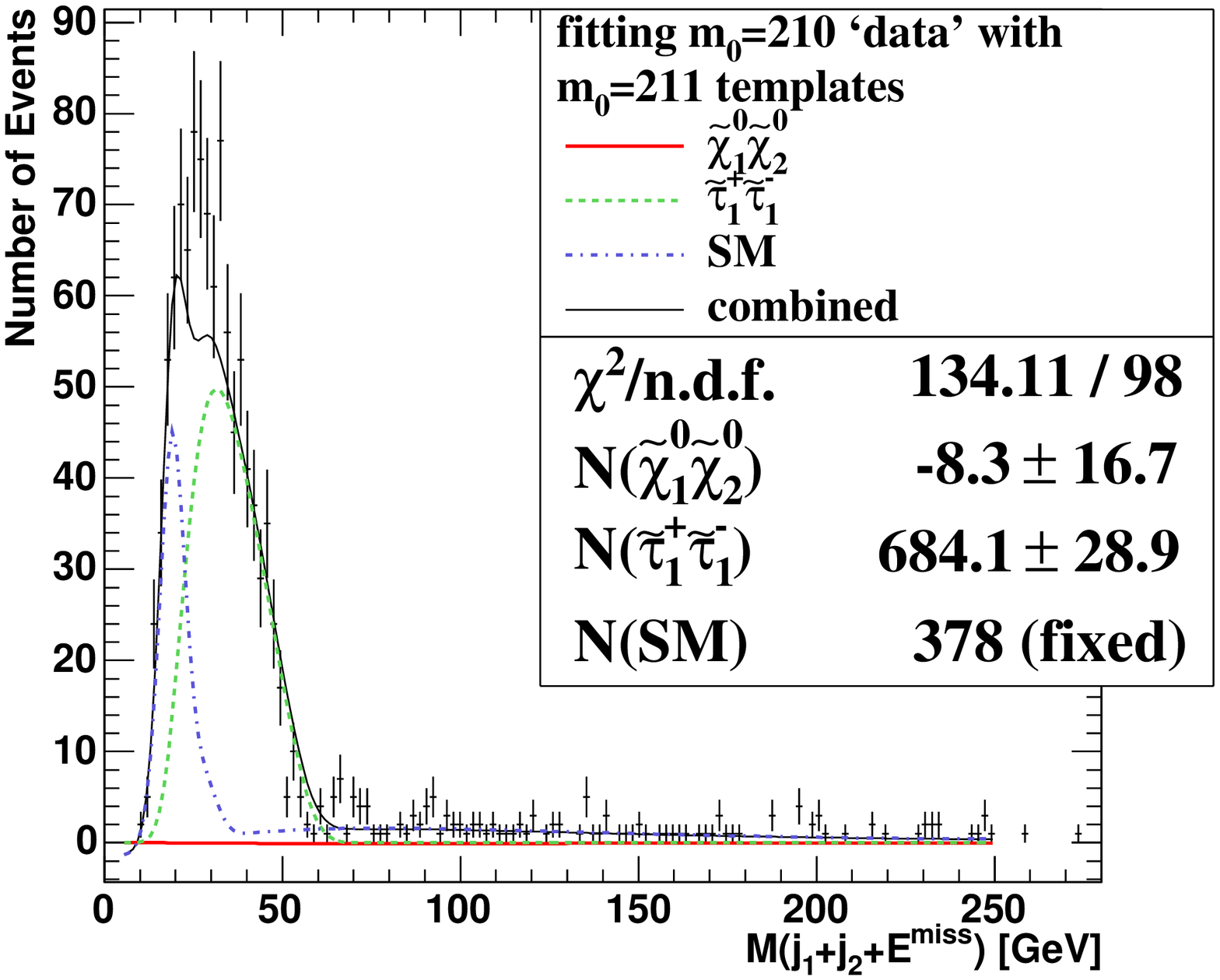}}
\caption{Example of fitting a MC sample containing 
SM and SUSY ($m_0$ = 210 GeV, $m_{1/2}$ = 360 GeV) events equivalent to 
500\ \invfb\ 
to two \meff\ templates
for SM+SUSY ($m_0$ = 210 or 211 GeV, $m_{1/2}$ = 360 GeV). 
A 2\degr\ VFD is assumed.
The value of $\chi^{2}$/n.d.f. is minimum when the events from the same
SUSY parameter are in the 500 \invfb\ sample. 
A $\gamma\gamma$ contribution (a narrow distribution around 20 GeV)
is apparent.
The fitting is similar for 1\degr\ VFD,
except the $\gamma\gamma$ contribution is substantially reduced.}
\label{fig8}
\end{figure}

The measurement of a small $\Delta M$ value is crucial since it would be a key evidence 
of the existence of
the $\stau$-$\schionezero$ co-annihilation. 
We propose the variable \meff\
as a key discriminator of the signal events
from its background events.
We first generate the high statistics MC samples for the SM and various SUSY events 
(by changing the $m_0$ value) and prepare the templates of the \meff\ distributions
for the SM, $\schionezero\schitwozero$, and $\stauonep\stauonem$ events.
Figure~\ref{fig8} (without the data points for 500\ \invfb) shows examples of such templates
for two $m_0$ values for
a 2\degr\ VFD in the RH polarization case. 
The SM cross section is fitted by a blue line, the stau pair by
a green line and the neutralino pair by a red line. 
The stau pair production
 peak separates from the SM background  as $\Delta M$ increases. 
This is because for 
smaller $\Delta M$, 
the two $\tau$ signal appears like the $\tau$'s 
from the two photon background and consequently
this
region requires a VFD coverage down to 1\degr. 
From Figure~\ref{fig8}
 we also find
that the stau pair production cross section can be measured upto an accuracy of
$\pm$ 4\% for Point 2.

\begin{figure}\vspace{-2cm}
\centerline{ \DESepsf(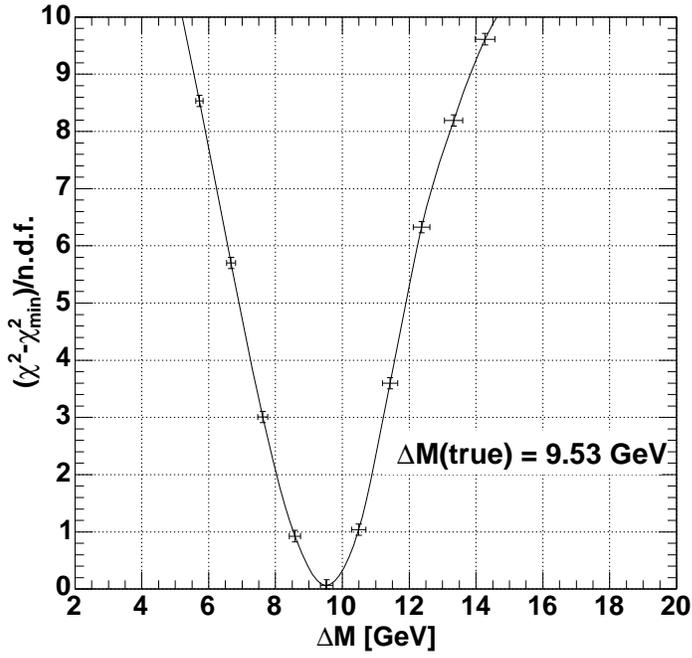 width 10 cm) }
\caption {\label{fig9}  $\chi^2$-$\chi^2_{\rm min}$  
of fitting the 500 \invfb\ sample 
for SUSY Point 2 ($m_0$ = 210 GeV, $m_{1/2}$ = 360 GeV)
with the high statistics templates  
is plotted as a function of $\Delta M$.} 
\end{figure}

Since the data of 500\ \invfb\ of luminosity will be generated in the initial run for a few
years, we then generate the MC samples equivalent to 500\ \invfb\ of luminosity
for particular $\Delta M$ values and
fit them with the template functions generated for high statistics sample. 
The black lines in Figure~\ref{fig8} shows the fitting of the  500 \invfb\ MC samples for point 2 with
the templates of two different $m_0$ values of 210 and 211 GeV. 
(Other parameters are kept at the same values as before.) We then compare the $\chi^2$ for
these fits.
We find that the $\chi^2$ for these fits is minimum for the $m_0$ = 210 GeV case. 
We use the range of $m_0$ = 203-220 GeV and try to fit the 500 \invfb\
MC sample for point 2 and determine the $\chi^2$ for all these
different points. 
We plot the $\chi^2$ of these fits in Figure~\ref{fig9}
 and find that 1$\sigma$ in the $\chi^2$ corresponds to  $9.5\pm 1$ GeV. 
The true value of $\Delta M$ for the point 2 is 9.53 GeV.
We repeat the same study for different $\stauone$ masses 
i.e. for different $\Delta M$. 
For lower $\Delta M$($\sim$ 5 GeV), 
we need to use a VFD of $1^\circ$. 
The accuracy of mass determination for
two different VFDs is summarized
in Table~\ref{table:dM_accuracy}, 
showing 
the uncertainties are at a level of 10\%, except  
for $\Delta M \sim$ 5 GeV
where it is 20\% and we have ~100 $\stauonep\stauonem$ events.

\begin{figure}\vspace*{1cm}
\centerline{\epsfxsize=10cm\epsfysize=10.0cm
\epsfbox{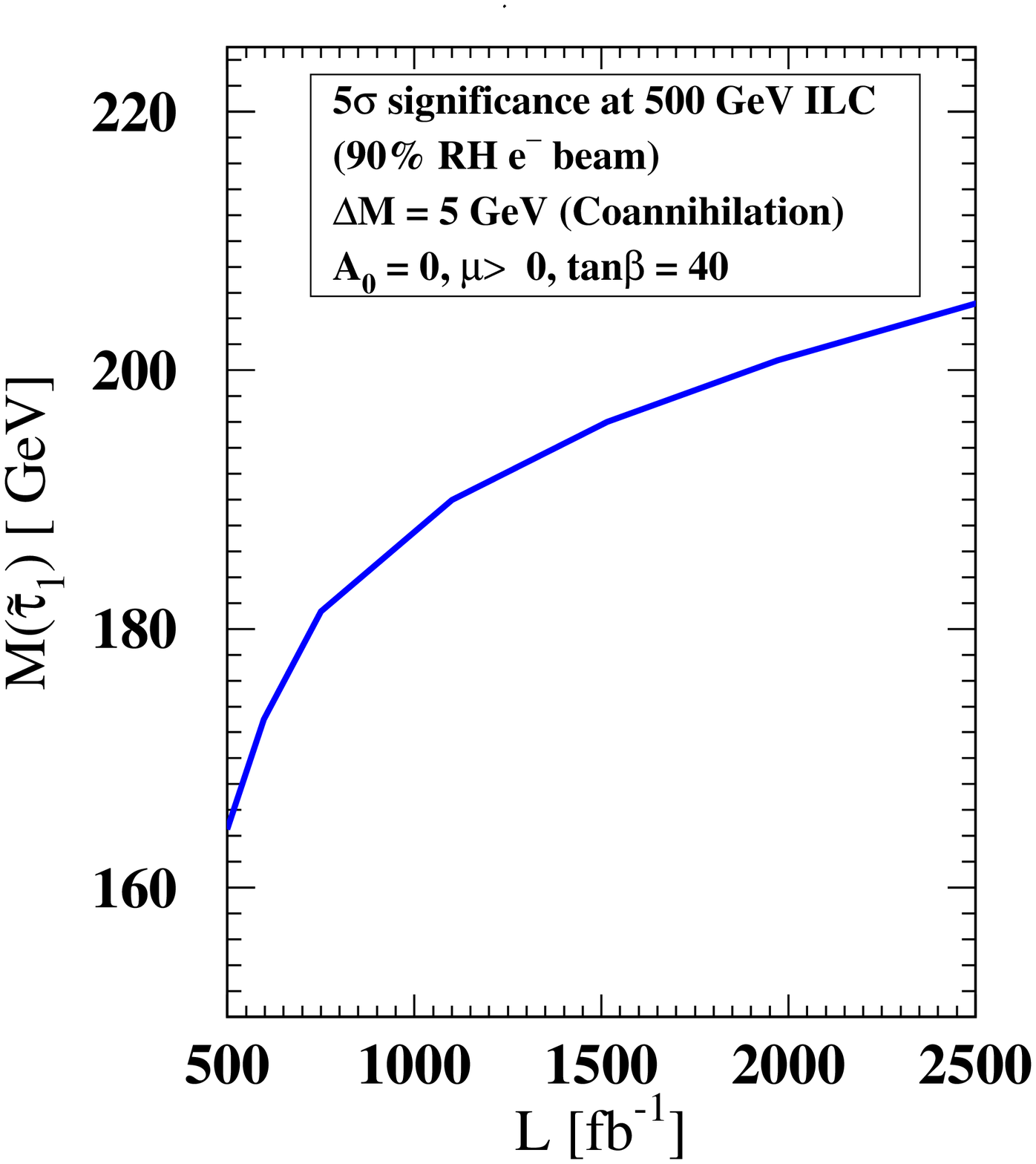}}
\caption{The $\stauone$ mass reach with $\Delta M= 5$ GeV as a function of  luminosity
 for a 5$\sigma$ discovery  with at least 100 events.}
\label{fig10}
\end{figure}

Figure~\ref{fig10} illustrates  the $\stauone$ mass reach 
as a function of  luminosity for a 5$\sigma$
discovery
with at least 100 events 
for $\Delta M\sim$ 5 GeV,
 where
the $\Delta M$ would be determined to 20\% or better.
We find that 164 GeV and 205 GeV $\stauone$ masses
to be $5\sigma$ reach and 20\% (or better) uncertainty
in $\Delta M$  measurement 
with 500 and  2500 \invfb\
at a 500 GeV ILC.

\begin{table}[t]
\caption{Accuracy of the determination of $\Delta M$ for different VFDs.}
\label{table:dM_accuracy}
\begin{center}
\begin{tabular}{c c| c| c c  }
\hline
\hline 
 &	& $N_{\stauonep\stauonem}$ &
	\multicolumn{2}{c}{$\Delta M$(``500 \invfb'' experiment)} \\ \cline{4-5} 
$m_0$ & $\Delta M$ & (500 \invfb) & 2$^\circ$ VFD  & 1$^\circ$ VFD \\
\hline
205 & 4.76  & 122 & Not Determined 	& 4.7$^{+1.0}_{-1.0}$\\
210 & 9.53  & 787 & 9.5$^{+1.1}_{-1.0}$	& 9.5$^{+1.0}_{-1.0}$\\
213 & 12.4 & 1027 & 12.5$^{+1.4}_{-1.4}$ & 12.5$^{+1.1}_{-1.4}$\\
215 & 14.3 & 1138 & 14.5$^{+1.1}_{-1.4}$ & 14.5$^{+1.1}_{-1.4}$\\
\hline
\hline
\end{tabular}
\end{center}\end{table}

\section{Conclusion}

We have probed the mSUGRA and SM signals 
in the $\stau$-$\schionezero$ co-annihilation region at a
500 GeV ILC with 500 \invfb of luminosity. 
In this region, 
the mass difference $\Delta M$ between the $\stauone$ and the $\schionezero$
would typically be 5-15 GeV for a large range of $m_{1/2}$. 
This small mass difference
would produce very low energy taus in the final state.  
The dominant SM background would be the two-photon process.
With RH $e^-$ beams
our study has focused on the
$\staup\staum$ production because
it allowed us to reach large $m_{1/2}$ values
in the allowed parameter space.
We proposed the invariant mass of two tau jets
and missing energy variable, $M(j_1,j_2,\menergy)$,
to determine the mass difference and
found the accuracy would be at a level of 10\% 
using a 1\degr\ (or 2\degr) VFD
except for $\Delta M$ = 4.76 GeV.
For $\Delta M \simeq$  5 GeV, 
a 1\degr\ VFD would be crucial
to suppress the two-photon background
and the accuracy there would be
about 20\% with approximately 100 signal events.
We also calculated 
the discovery significance of this region 
and determined the $5\sigma$ reach in $m_{1/2}$ for
500 \invfb of luminosity.

\section{Acknowledgments}
This work is supported in part by a NSF Grant
PHY-0101015, in part by  NSERC 
of Canada and in part by a DOE Grant DE-FG02-95ER40917.

\newpage



\begin{thebibliography}{99}

\bibitem{tesla} TESLA: Technical design report, part 4: A detector for
TESLA, T. Behnke \etal, DESY-2001-011, DESY-2001-011D,
DESY-TESLA-2001-23, DESY-TESLA-FEL-2001-05, ECFA-2001-209, March 2001. 
In a recent upgrade of the final focus setup, the forward calorimeter,
low angle tagger or LAT,
covers an angle down to 3.2 mrad.

\bibitem{sugra1} A.H. Chamseddine, R. Arnowitt, and P. Nath,
Phys. Rev. Lett. {\bf 49} (1982) {970}. 

\bibitem{sugra2} R.~Barbieri, S.~Ferrara, and C.~A.~Savoy,
Phys. Lett. {\bf B119} (1982) {343}; 
L. Hall, J. Lykken, and S. Weinberg, Phys. Rev. {\bf D27} (1983) {2359}; 
P. Nath, R. Arnowitt, and A.H. Chamseddine,
Nucl. Phys. {\bf B227} {(1983)} {121}.

\bibitem{nilles}
For a review, P. Nilles, { Phys. Rept.} {\bf 110} {(1984)} 1.

\bibitem{dark}J. Ellis, K. Olive, Y. Santoso, and V. Spanos,  
Phys. Lett. {\bf B565} (2003) 176; 
R. Arnowitt, B. Dutta, T. Kamon, and V. Khotilovich, hep-ph/0308159; 
H. Baer, C. Balazs, A. Belyaev, 
T. Krupovnickas, and X. Tata, JHEP {\bf 0306} (2003) 054; 
B. Lahanas and D.V. Nanopoulos, Phys. Lett. {\bf B568} (2003) 55;
 U. Chattopadhyay, A. Corsetti, and P. Nath, Phys. Rev. {\bf D68} (2003) 035005.
\bibitem{exp1} H. Baer, T. Krupovnickas, and X. Tata, JHEP {\bf 0406} (2004) 061.
\bibitem{exp2} P. Bambade, M. Berggren, F. Richard, and Z. Zhang, hep-ph/0406010.
\bibitem{higgs1} See, for example, P.~Igo-Kemenes, LEPC meeting,   
(http://lephiggs.web.cern. ch/LEPHIGGS/talks/index.html).

\bibitem{bsgamma} M. Alam \etal, Phys. Rev. Lett. {\bf 74} {(1995)} {2885}. 

\bibitem{sp} WMAP Collaboration, D.N Spergel \etal,
Astrophys. J. Suppl. {\bf 148} (2003) 175.

\bibitem{aleph} ALEPH collaboration, ALEPH-CONF-2001-009.

\bibitem{BNL} Muon $g-2$ Collaboration, G. Bennett \etal, 
Phys. Rev. Lett. {\bf 92} (2004) 161802.

\bibitem{dav} M. Davier, hep-ex/0312065.

\bibitem{hag} K. Hagiwara, A. Martin, D. Nomura, and T. Teubner,  
Phys. Rev. {\bf D69} (2004) 093003.

\bibitem{mar}W. Marciano, hep-ph/0411179.

\bibitem{rattazi} R. Rattazi and U. Sarid, Phys. Rev. {\bf 53} {(1996)} {1553};
 M. Carena, M. Olechowski, S. Pokorski, and C. Wagner, Nucl. Phys. 
 {\bf 426} {(1994)} {269}.

\bibitem{degrassi}G. Degrassi, P. Gambino, and G. Giudice, 
{JHEP} {\bf 0012} {(2000)} {009}; M. Carena, D. Garcia, U. Nierste,
and C. Wagner,
Phys. Lett. {\bf B499} {(2001)} {141};
G. D'Ambrosio, G. Giudice, G. Isidori, and A.Strumia, hep-ph/0207036;
A. Buras, P. Chankowski, J. Rosiek, and L. Slawianowska, hep-ph/0210145.

\bibitem{bdutta} R. Arnowitt, B. Dutta, and Y. Santoso, hep-ph/0010244; 
hep-ph/0101020; Nucl. Phys. {\bf B606} {(2001)} {59}; 
J Ellis, T. Falk, G. Ganis, K. Olive, and M. Srednicki,
Phys. Lett. {\bf B570} {(2001)} {236};  
J. Ellis, T. Falk, and K. Olive, Phys. Lett. {\bf B444}  {(1998)} {367};
J. Ellis, T. Falk, K. Olive, and M. Srednicki, {Astro. Phys.} {\bf 13}  {(2000)} {181};
Erratum-ibid.{ \bf 15} {2001} {413}; 
M. Gomez, and J. Vergados, Phys. Lett. {\bf B512} {(2001)} {252}; 
M. Gomez, G. Lazarides, and C. Pallis, Phys. Rev. {\bf D61} {(2000)} {123512}; 
Phys. Lett. {\bf B487} {(2000)} {313};
 L. Roszkowski, R. Austri, and T. Nihei, {JHEP} {\bf 0108} {(2001)} {024}; 
 A. Lahanas, D.V. Nanopoulos, and V. Spanos, Phys. Lett. {\bf B518} {(2001)} {518}.

\bibitem{isajet} F. Paige, S. Protopescu, H. Baer, and X. Tata, hep-ph/0312045.
We use {\tt ISAJET} version 7.69.

\bibitem{wphact}
E. Accomando and A. Ballestrero, Comput. Phys. Commun. {\bf 99} (1997) 270;
E. Accomando, A. Ballestrero, and E. Maina, Comput. Phys. Commun. {\bf 150} (2003) 166.
We use {\tt WPHACT} version 2.02pol.

\bibitem{tauola}
S. Jadach and J. H. K\"uhn, {Comput. Phys. Comm.} {\bf 64} {(1991)} {275};
M. Je\.zabek, Z. Was, S. Jadach,
and J. H. K\"uhn, {Comput. Phys. Comm.} {\bf 70} {(1992)} {69};
M. Je\.zabek, Z. Was,
S. Jadach, and J. H. K\"uhn, {Comput. Phys. Comm.} {\bf 76} {(1993)} {361}.
We use {\tt TAUOLA} version 2.6.

\bibitem{LCD} LCD Root Package  version 3.5 with LD Mar01 detector 
parametrization. See, for example, 
T.~Abe and M.~Iwasaki,
in Proc. of the APS/DPF/DPB Summer Study on the Future of Particle Physics 
(Snowmass 2001), ed. N.~Graf,
eConf {\bf C010630}, E3045 (2001)
[hep-ex/0110068].

\bibitem{jade} JADE collaboration, W. Bartl \etal, {Z. Phys} {\bf C33} {(1986)} {23};
 S. Bethke \etal, Phys. Lett. {\bf B213} {(1988)} {235}.
\end{thebibliography}
\end{document}